\documentclass{svproc}
\usepackage{url}

\usepackage{tabularx}
\usepackage{booktabs}

\usepackage{graphicx}
\usepackage{hyperref}
\usepackage[sorting=none,url=false]{biblatex}
\usepackage[misc,geometry]{ifsym}
\newcommand{\corrAuthor}{$^{\textrm{\Letter}}$}

\newcommand{\multicell}[2][t]{\begin{tabular}[#1]{@{}l@{}}#2\end{tabular}} % allow linebreaks in cell

\usepackage{tikz}
\newcommand\copyrighttext{%
  \footnotesize The final authenticated publication is available online at \newline \url{https://doi.org/10.1007/978-3-030-67084-9_8}.
  }
\newcommand\copyrightnotice{%
\begin{tikzpicture}[remember picture,overlay]
\node[anchor=south,yshift=10pt] at (current page.south) {\fbox{\parbox{\dimexpr\textwidth-\fboxsep-\fboxrule\relax}{\copyrighttext}}};
\end{tikzpicture}%
}

\begin{document}
\mainmatter
\title{Experience vs Data: A Case for More Data-informed Retrospective Activities}
\titlerunning{Experience vs Data}  % abbreviated title (for running head)
\author{Christoph Matthies$^{[0000-0002-6612-5055]}$\corrAuthor, Franziska Dobrigkeit$^{[0000-0001-9039-8777]}$}
\authorrunning{Matthies, Dobrigkeit} % abbreviated author list (for running head)
% \authorrunning{Blinded} % abbreviated author list (for running head)
\institute{Hasso Plattner Institute\\ University of Potsdam, Germany\\
\email{christoph.matthies@hpi.de, franziska.dobrigkeit@hpi.de}\\
}

\maketitle              % typeset the title of the contribution

% Render copyright notice
\copyrightnotice

\begin{abstract}
Effective Retrospective meetings are vital for ensuring productive development processes because they provide the means for Agile software development teams to discuss and decide on future improvements of their collaboration. 
Retrospective agendas often include activities that encourage sharing ideas and motivate participants to discuss possible improvements.
The outcomes of these activities steer the future directions of team dynamics and influence team happiness.
However, few empirical evaluations of Retrospective activities are currently available. Additionally, most activities rely on team members experiences and neglect to take existing project data into account.
With this paper we want to make a case for data-driven decision-making principles, which have largely been adopted in other business areas. Towards this goal we review existing retrospective activities and highlight activities that already use project data as well as activities that could be augmented to take advantage of additional, more subjective data sources.
We conclude that data-driven decision-making principles, are advantageous, and yet underused, in modern Agile software development. Making use of project data in retrospective activities would strengthen this principle and is a viable approach as such data can support the teams in making decisions on process improvement.
\end{abstract}
\section{Introduction}
\begin{figure}[htb]
    \centering
	\includegraphics[width=0.85\linewidth]{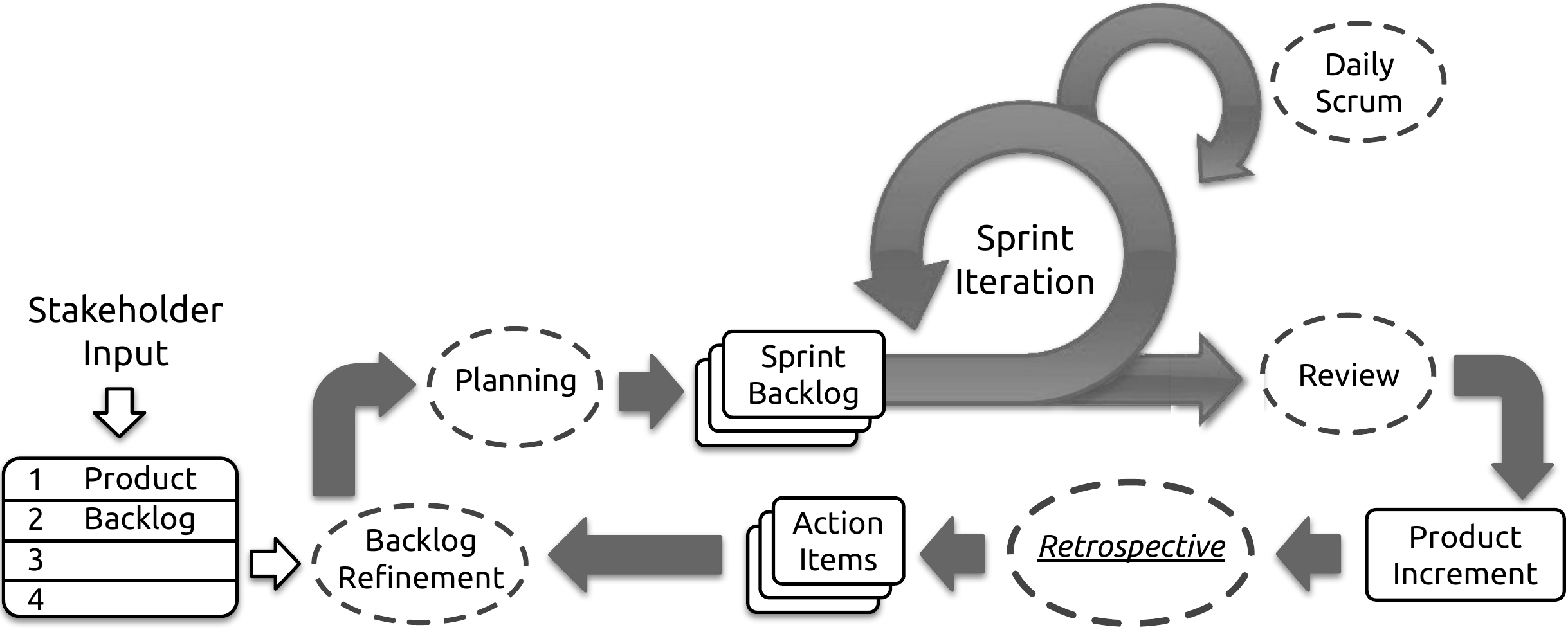}
	\caption{Prototypical flow through the Scrum process, based on~\cite{Matthies2020HICSS}. Process meetings are represented by circles, process artifacts and outcomes as squares. Scrum's process improvement meeting, the Retrospective, is highlighted.}
	\label{fig:process}
\end{figure}

Agile development methods, particularly Scrum, which focus on managing the collaboration of self-organizing, cross-functional teams working in iterations~\cite{Schwaber2017}, have become standards in industry settings.
The most recent survey of Agile industry practitioners by \textit{Digital.ai}\footnote{formerly \textit{CollabNet VersionOne}}, conducted between August and December 2019, showed that Scrum continued to be the most widely-practiced Agile method:
75\% of respondents employed Scrum or a Scrum hybrid~\cite{StateOfAgile2020}.
In the survey, which included 1,121 full responses, both of the top two Agile techniques employed in organizations were focused on communication and gathering feedback: the Daily Standup (85\%) and Retrospective meetings (81\%).
The importance of these meetings was also stated in a previous, similar survey by the \textit{Scrum Alliance}.
A vast majority of respondents (81\%) to this 2018 survey stated that their teams held a Retrospective meeting at the end of every Sprint, while 87\% used Daily Scrum meetings~\cite{ScrumAlliance2018}.
A prototypical, generalized flow through the Scrum method, depicting the different prescribed meetings and process artifacts, i.e. the context of Retrospectives, is represented in Figure~\ref{fig:process}.

In this research, we focus on the popular Retrospective meeting, which forms the core practice of process improvement approaches in the Scrum method, and the activities that are used in it.
Retrospectives are a realization of the ``inspect and adapt'' principle~\cite{Schwaber2017} of Agile software development methods~\cite{Andriyani2017}.

\subsection{Retrospective Meetings}
Recent research has pointed to Retrospective meetings as crucial infrastructure in Scrum~\cite{dingsoyr2018}.
Similarly, Retrospectives have also been recognized as one of the most important aspects of Agile development methods by practitioners~\cite{Kniberg2015}.
The seminal work on Scrum, the \emph{Scrum Guide}, defines the goal of Retrospectives to ascertain ``how the last Sprint went'' regarding both people doing the work, their relationships, the employed process, and the used tools~\cite{Schwaber2017}.
As such, Retrospectives cover improvements of both technical and social/collaboration aspects.
Teams are meant to improve their modes of collaboration and teamwork, thereby also increasing the enjoyment in future development iterations~\cite{Derby2006}.
The Scrum framework prescribes Retrospectives at the end of each completed iteration.
Teams are meant to generate a list of improvement opportunities, i.e. ``action items''~\cite{Derby2006}, to be tackled in the next iteration.
Retrospective meetings focus less on the quality of the produced product increment, but more on how it was produced and how that process can be made smoother and more enjoyable for all involved parties in the next iteration.

While Scrum is a prescriptive process framework, suggesting concrete meetings, roles, and process outcomes, the Scrum Guide also points out: ``Specific tactics for using the Scrum framework vary and are described elsewhere''~\cite{Schwaber2017}.
For Retrospectives, this means that while the meeting's goal of identifying improvement opportunities is clear, the concrete steps that teams should follow are not and are up to the individual, self-organizing Scrum teams~\cite{Matthies2016c,Przybyek2017}.
One of the easiest and most effective ways to generate the types of process insights that Retrospectives require is by relying on those most familiar with the teams' executed processes: team members themselves.
Their views and perceptions of the previous, completed development iteration are, by definition, deeply relevant as inputs for process improvement activities.
Furthermore, these data points are collectible with minimal overhead, e.g. by facilitating a brainstorming session in a Retrospective meeting, and they are strongly related to team satisfaction~\cite{caroli2015}.

\subsection{Data Sources used in Retrospective Meetings}
Most of the data that forms the basis of improvement decisions in current Retrospectives is, at present, based on the easily collectible perceptions of team members.
However, modern software development practices and the continuing trend of more automated and integrated development tools have opened another avenue for accessing information on teams' executed process: their \emph{project data}~\cite{Matthies2020CHASE,Zaitsev2020,Wohlrab2020}.
This project data includes information from systems used for such diverse purposes as version control (what was changed, why, when?), communication (what are other working on?), code review (feedback on changes), software builds (what is the testing status?), or static analysis (are standards met?).
The data is already available, as modern software engineers continuously document their actions as part of their regular work~\cite{Ying2005,Matthies2016b}.
The development processes of teams, their successes as well as their challenges, are ``inscribed'' into the produced software artifacts~\cite{de2005seeking}.
This type of information, which can be used in Retrospectives, in addition to the subjective assessments of team members, has been identified as ``a gold-mine of actionable information''~\cite{Guo2016}.
More comprehensive, thorough insights into teams' process states, drawn from activities that make use of both project data and team members' perceptions, can lead to even better results in Retrospectives~\cite{Derby2006}.

\subsection{Research Goals}
In this research, we focus on the integration of project data sources into Agile Retrospective meetings.
In particular, we investigate to which extent project data analyses are already provided for in Retrospective activities and how more of them could benefit from data-informed approaches in the future.
We provide an overview of popular activities and review the types of data being employed to identify action items, i.e. possible improvements.
We highlight those activities that already rely on software project data in their current descriptions as well as those that could be augmented to take advantage of the information provided by project data sources.
We argue that the principles of data-driven decision making, which have already been adopted in many business areas~\cite{Matthies2019ICSOFT}, are suitable and conducive, yet underused, in the context of modern Agile process improvement.

\section{Retrospective Activities}
The core concept of Retrospectives is not unique to Scrum. 
These types of meetings, focusing on the improvement of executed process and collaboration strategies, have been employed since before Agile methods became popular.
Similarly, team activities or ``games'' that meeting participants can play to keep sessions interesting and fresh have been used in Retrospectives since their inception~\cite{kerth2000}.
These, usually time-boxed, activities are interactive and designed to encourage reflection and the exchange of ideas in teams.
Derby and Larsen describe the purpose of Retrospective activities as to ``help your team think together''~\cite{Derby2006}.
Retrospective games have been shown to improve participants’ creativity, involvement, and communication as well as make team members more comfortable participating in discussions~\cite{Przybyek2017}.
The core idea is that meeting participants already have much of the information and knowledge needed for future process improvements, but a catalyst is needed to start the conversation.

In 2000, Norman L. Kerth published a collection of Retrospective activities~\cite{kerth2000}.
Additional collections were published in the following years by different practitioners as well as researchers~\cite{hohmann2006,kua2013,goncalves2014,krivitsky2015}.
Table~\ref{table:activity_src} presents an overview of the literature containing collections of Retrospective activities.
\begin{table}[htb]
    % \ra{1.2}
    \centering
    \caption{Sources of Retrospective activities in literature.}
    \label{table:activity_src}
    \begin{tabular}{lrl}
        \toprule
        \textbf{Year} & \textbf{Reference} & \textbf{Name of Reference}\\
        \midrule
        2006 & \cite{Derby2006} & Agile Retrospectives - Making Good Teams Great \\
        2006 & \cite{hohmann2006} & Innovation Games \\
        2013 & \cite{kua2013} & The Retrospective Handbook \\
        % 2013 & Project Retrospectives: A Handbook for Team Reviews~\cite{kerth2013} \\
        2014 & \cite{goncalves2014} & Getting value out of Agile Retrospectives \\
        2015 & \cite{krivitsky2015} & Agile Retrospective Kickstarter \\
        2015 & \cite{caroli2015} & Fun Retrospectives \\
        2018 & \cite{Baldauf2018} & Retromat: Run great agile retrospectives! \\
        \bottomrule
    \end{tabular}
\end{table}

A generalized meeting agenda for Retrospective was proposed in 2006 by Derby and Larsen. 
It features five consecutive phases: (i) \textit{set the stage} (define the meeting goal and giving participants time to ``arrive''), (ii) \textit{gather data} (create a shared pool of information), (iii) \textit{generate insight} (explore why things happened, identify patterns within the gathered data), (iv) \textit{decide what to do} (create action plans for select issues), and (v) \textit{close} (focus on appreciations and future Retrospective improvements)~\cite{Derby2006}.
This plan has since established itself and has been accepted by other authors~\cite{Baldauf2018}.
Retrospective activities remain an open area of investigation and continued learning, with current research further exploring the field~\cite{jovanovic2016,loeffler2017,Matthies2020HICSS}.

While research articles and books offer extensive collection efforts regarding Retrospective activities, Agile practitioners rely on up-to-date web resources in their daily work, rather than regularly keeping up with research literature~\cite{loeffler2017,Beecham2014}.
The \emph{Retromat}\footnote{available at \url{https://retromat.org}}~\cite{Baldauf2018} is a popular, comprehensive and often referenced~\cite{northwood2018,loeffler2017,Kniberg2015}, online repository of Retrospective activities for meeting agendas.
It currently contains 140 different activities for five Retrospective meeting phases~\footnote{\url{https://retromat.org/blog/history-of-retromat/}}.

\section{Review of Retrospective Activities}
As the Retromat repository represents the currently best updated, most complete list of Retrospective activities in use by practitioners~\cite{northwood2018,loeffler2017,Stalesen2015}, we employ its database as the foundation of our review.
Our research plan contains the following steps:
\begin{itemize}
    \item Extract activities that provide or generate inputs for discussion in Retrospectives
    \item Identify the specific data points being collected
    \item Categorize data points by their origin
    \item Study those activities in detail which already (or are close to) taking project data into account
\end{itemize}

\subsection{Activity Extraction}
The Retromat, following Derby and Larsen's established model~\cite{Derby2006}, features activities and games for the five Retrospective phases~\footnote{\url{https://retromat.org/blog/what-is-a-retrospective/}} \textit{set the stage}, \textit{gather data}, \textit{generate insight}, \textit{decide what to do} and \textit{close the Retrospective}.
As this research focuses on the types of gathered inputs employed for meetings, we initially collected all activities classified by the Retromat as suitable for the \emph{gather data} phase.
This meeting phase aims to help participants remember and reflect and is aimed at collecting the details of the last iteration, in order to establish a shared understanding within the team.
We extracted 35 activities intended for the \emph{gather data} from the Retromat repository.
These activities are listed in Table~\ref{table:activities} in the Appendix.

Additionally, we reviewed the Retromat activities prescribed for all other phases to ensure that we did not miss any activities that gathered data as part of their proceedings.
These could have been classified under different phases, as data gathering and analysis steps are often intertwined, or because the activity's main focus is broader than data collection.
This step yielded an additional four activities that, at least partly, base their procedures on collected data: ``3 for 1 - Opening'' (assessments of iteration results and number of communications), ``Last Retro's Actions Table'' (collecting assessments of previous action items), ``Who said it?'' (collecting memorable quotes), and ``Snow Mountain'' (using the Scrum burndown chart)\footnote{\url{https://retromat.org/en/?id=70-84-106-118}}.
The first three of these were classified in the Retromat under the \emph{set the stage} phase, the last as \emph{generate insights}.

\subsection{Identification of Retrospective Inputs}
We analyzed the textual descriptions provided within the Retromat collection for each of the extracted activities of the previous research step.
We manually tagged each of the activities with labels regarding the specific data points that are collected and used as inputs for the following actions.
Many activity descriptions featured subsequent aggregation and synthesis actions, e.g. dot-voting or clustering, from which we abstracted.
The generated short data labels describe the specific outcomes of the initial data acquisition within activities.
Examples include ``numerical ratings of performed meetings'',  ``notes on what team members wish the team would learn'', or ``collection of all user stories handled during the iteration''.
Multiple activity descriptions contained mentions of physical representations of collected data points, which we generalized.
For example, we consider ``index cards'' and ``sticky notes'' filled by meeting participants with their ideas to be instances of the more general ``notes''.
The results of this tagging step are shown in Table~\ref{table:type_input}.

\begin{table}[!htbp]
    \centering
    \caption{Overview of the types of inputs employed in the selected Retrospective activities. Activities above the divide are part of the \emph{gather data} phase, those below are included after reviewing the activities of other phases. The categories of activities that gather data through specific prompts are italicized.}
    \label{table:type_input}
    \begin{tabularx}{\columnwidth}{@{}llX@{}}
        \toprule
        \textbf{Shortened name} & \textbf{\#} & \textbf{Type of activity input (regarding last iteration)} \\
        \midrule
        & & \\
\multicolumn{3}{l}{Activities from the \emph{gather data} Retrospective phase} \\
\midrule
\textit{Timeline} & 4 & List of memorable/personally significant events \\
\textit{Analyze Stories} & 5 & Collection of user stories handled during the iteration \\
\textit{Like to like} & 6 & Notes on things to \emph{start doing}, \emph{keep doing} and \textit{stop doing} \\
\textit{Mad Sad Glad} & 7 & Notes on events when team members felt \textit{mad}, \textit{sad} or \textit{glad} \\
\textit{Speedboat/Sailboat} & 19 & Notes on what drove the team \textit{forward} \& what \textit{kept it back} \\
\textit{Proud \& Sorry} & 33 & Notes of instances of \textit{proud} and \textit{sorry} moments \\
\textit{Self-Assessment} & 35 & Assessments of team state regarding Agile checklist items \\
\textit{Mailbox} & 47 & Reports of events or ideas collected during the iteration \\
\textit{Lean Coffee} & 51 & List of topics team members wish to be discussed \\
\textit{Story Oscars} & 54 & Physical representations of completed user stories \\
\textit{Expectations} & 62 & Text on what team members expect from each other \\
\textit{Quartering} & 64 & Collection of everything the team did during iteration \\
\textit{Appreciative Inquiry} & 65 & Answers to positive questions, e.g. best thing that happened \\
\textit{Unspeakable} & 75 & Text on the biggest unspoken taboo in the company \\
\textit{4 Ls} & 78 & Notes on what was \textit{loved}, \textit{learned}, \textit{lacked} \& \textit{longed for} \\
\textit{Value Streams} & 79 & Drawing of a value stream map of a user story \\
\textit{Repeat \& Avoid} & 80 & Notes on what practices to \textit{avoid} and which to \textit{repeat} \\
\textit{Comm. Lines} & 86 & Visualization of the ways information flows in the process \\
\textit{Satisfaction Hist.} & 87 & Numerical (1-5) ratings of performed meetings \\
\textit{Retro Wedding} & 89 & Notes on categories something \textit{old}, \textit{new}, \textit{borrowed} \& \textit{blue} \\
\textit{Shaping Words} & 93 & Short stories on iteration, including a 'shaping word' \\
\textit{\#tweetmysprint} & 97 & Short texts/tweets commenting on the iteration \\
\textit{Laundry Day} & 98 & Notes on \textit{clean} (clear) \& \textit{dirty} (unclear/confusing) items\\
\textit{Movie Critic} & 110 & Notes on movie critic-style categories: \textit{Genre}, \textit{Theme}, \textit{Twist}, \textit{Ending}, \textit{Expected?}, \textit{Highlight}, \textit{Recommend?} \\
\textit{Genie in a Bottle} & 116 & Notes on 3 wishes: for \textit{yourself}, \textit{your team} and \textit{all people} \\
\textit{Hit the Headlines} & 119 & Short headlines on newsworthy aspects of the iteration \\
\textit{Good, Bad \& Ugly} & 121 & Notes on categories \textit{good}, \textit{bad} \& \textit{ugly} concerning the iteration \\
\textit{Focus Principle} & 123 & Assessments on relative importance of Agile Manifesto principles\\
\textit{I like, I wish} & 126 & Notes on \textit{likes} and \textit{wishes} concerning the iteration \\
\textit{Delay Display} & 127 & Notes on team \textit{destination}, \textit{delay} \& \textit{announcement} \\
\textit{Learning Wish List} & 128 & Text on what team members wish the team would learn \\
\textit{\multicell{Tell me something\\I don’t know}} & 133 & Facts and questions, in game show fashion, on something that only one team member knows and most others do not \\
\textit{Avoid Waste} & 135 & Notes on the \textit{7 categories of waste} in the process \\
\textit{Dare, Care, Share} & 137 & Notes on \textit{bold wishes}, \textit{worries} \& \textit{feedback/news} \\
\textit{Room Service} & 139 & Notes on the prompts \textit{Our work space helps me/us...} and \textit{Our work space makes it hard to...} \\
& & \\
\multicolumn{3}{l}{Activities from phases \emph{set the stage} and \emph{generate insights}} \\
\midrule
\textit{3 for 1} & 70 & Points in coordinate plane of satisfaction with results and communication \\
\textit{Retro Actions Table} & 84 & List of last Retrospectives action items \\
\textit{Who said it?} & 106 & Quotes collected from project artifacts \\
\textit{Snow Mountain} & 118 & Burndown chart of problematic Sprint \\
        \bottomrule
    \end{tabularx}
\end{table}

\subsection{Classification of Retrospective Data Sources}
We categorized activities based on the origins of their data inputs, using the generated descriptions.
We distinguish whether the gathered data is (i) drawn solely from team members' perceptions (no mention/reliance on software project data), (ii) is directly extracted from project data sources, or (iii) is ambiguous, i.e. could be drawn from either source, depending on team context and interpretation.
We consider the term ``project data'' as an overarching collection of software artifacts.
We follow Fernández et al.'s definition of the term ``software artifacts''~\cite{Fernandez2018}, in that we consider them ``deliverables that are produced, modified, or used by a sequence of tasks that have value to a role''. 
These artifacts are often subject to quality assurance and version control and have a  specific type~\cite{Fernandez2018}.

The vast majority, i.e. 86\% (30 of 35), of proposed \emph{gather data} activities in the Retromat collection make no mention of software project data and do not take advantage of it.
It should be noted that most of these \emph{gather data} activities are very similar in terms of the type of collected data.
They tend to deal with team members' answers to varying prompts or imagined scenarios, aimed at starting discussions.
Examples of such prompts include ``mad, sad, glad``, ``start, stop, continue``, ``good, bad, ugly'' or ``proud and sorry''.
All of these activities are, by default, drawn from the individual perceptions and experiences of team members.

The nine activities that we identified in our review as featuring (possible) connections to development data---five from the \emph{gather data}, three from \emph{set the stage} and a single one from the \emph{generate insights} phase---are shown in Table~\ref{table:data_activities} and are discussed in the following two sections.

\begin{table}[!htbp]
    \centering
    \caption{Overview of Retromat activities not solely reliant on team members' perceptions. Activities which could be connected to project data, depending on how they are executed, are marked as \emph{Possible}.}
    \label{table:data_activities}
    \begin{tabularx}{\textwidth}{llXl}
        \toprule
        \multicell{\\\textbf{\#}} & \multicell{\textbf{Activity}\\\textbf{Name}} & \multicell{\textbf{Data used as (partial) input for the}\\\textbf{activity and subsequent steps}} & \multicell{\textbf{Project}\\\textbf{Data}}\\
        \midrule
        5 & Analyze Stories & Collection of all user stories handled during the iteration & \textit{Yes} \\
        54 & Story Oscars & Physical representation of all stories completed in the last iteration & \textit{Yes} \\
        84 & \multicell{Last Retro's Actions Table} & List of outcomes of the last Retrospective, i.e. action items/improvement plans & \textit{Yes} \\
        106 & Who said it? & Literal quotes of team members extracted from communication channels, e.g. emails, chat logs or ticket discussions & \textit{Yes} \\
        35 & Agile Self-Assessment & Assessments of team state regarding Agile checklist items & \textit{Possible} \\
        64 & \multicell{Quartering -\\Identify boring stories} & Collection of ``everything'' the team did in the last iteration & \textit{Possible} \\
        70 & 3 for 1 & Number of times team members coordinated in the last iteration & \textit{Possible} \\
        79 & Value Stream Mapping & Drawing of a value stream map concerning a particular user story & \textit{Possible} \\
        118 & Snow Mountain & The shape of the Scrum Burndown chart of a problematic iteration & \textit{Possible} \\
        \bottomrule
    \end{tabularx}
\end{table}

\section{Activities Already Reliant on Project Data}
Of the overall nine activities identified in this research that feature (possible) connections to project data, four make direct mentions of specific development artifacts in their descriptions on Retromat:

\begin{itemize}
    \item \emph{Analyze Stories}
    \item \emph{Story Oscars}
    \item \emph{Last Retro's Actions Table}
    \item \emph{Who said it?}
\end{itemize}

These are marked as \emph{Yes} regarding the use of project data in Table~\ref{table:activities}.
Of these four activities, two employ the user stories of the last iteration as inputs, which are analyzed and graded by meeting participants in the following steps.
The other two are concerned with the outcomes of the last Retrospective meeting and an extract of intra-team communications.
The user stories/work items of modern Agile teams are usually contained in an issue tracker system~\cite{Dimitrijevic2015} or can be acquired in printed form from a shared workspace or board~\cite{Kniberg2015}.
Persisting the outcomes of Retrospectives, i.e. making note of the resulting action items and documenting meeting notes, is a common practice of Agile processes~\cite{dingsoyr2018} and enables the tracking of progress towards these goals.
Furthermore, digital communication tools, e.g. bug reports, mailing lists, or online forums, and the artifacts that result from their usage form a core part of modern software development~\cite{Nazar2016}.
The fact that these project artifacts are already present and are produced as part of the regular tasks of modern software developers, means that they can be collected with minimal overhead~\cite{Ortu2015}.

The four Retrospective activities we identified in this review as already employing project data represent only a small fraction of the 140 overall activities included in the Retromat.
However, these are the activities that explicitly follow Derby and Larsen's principle of having the \emph{gather data} phase of Retrospective meetings ``start with the hard data''~\cite{Derby2006}.
The authors consider this ``hard data'' to include iteration events, collected metrics, and completed features or user stories.
They point out that while it ``may seem silly to gather data for an iteration that lasted a week or two'', being absent for a single day of a week-long iteration already results in missing 20\% of events.
As such, reflecting on the completed iteration through the lens of project data can ensure a more complete overview for all team members.
Furthermore, even when nothing was missed through absence, perceptions of iteration events vary between observers and different people exhibit different perspectives and understandings regarding the same occurrences~\cite{Derby2006}.
Lastly, by focusing on project data, in addition to the ``soft data'' usually employed, teams can optimize their Retrospective meetings.
The roles in teams tasked with facilitating Retrospectives are able to prepare the inputs for meeting activities beforehand, without relying on the presence of others.
Team members are able to focus their attention on interpreting data instead of trying to remember the details of the last iteration.
The time gained by reviewing, e.g. an already existing list of user stories rather than having to reconstruct it collaboratively, frees up more time for the actual Retrospective work of reflecting on process improvements using Retrospective activities.

\section{Towards Data-informed Retrospective Activities}
The activities that we identified, depending on interpretation and context, as having a \emph{possible} connection to project data, i.e. depending on concrete execution in teams, are ``Agile Self-Assessment'', ``Quartering - Identify boring stories'', ``3 for 1'', ``Value Stream Mapping'' and ``Snow Mountain'', see Table~\ref{table:data_activities}.
In the following paragraphs, we discuss these activities and their relations with software project data in detail.

\paragraph{Agile Self-Assessment} involves assessments of team members regarding the state of their own team, based on a checklist of items.
Depending on the employed checklist, these assessments might involve quantifiable measurements, e.g. ``time from pushing code changes until feedback from a test is received''\footnote{\url{https://finding-marbles.com/2011/09/30/assess-your-agile-engineering-practices/}} or can rely on entirely team members' perceptions, e.g. ``the team delivers what the business needs most''\footnote{\url{https://www.crisp.se/gratis-material-och-guider/scrum-checklist}}.
By switching to a checklist featuring measurements based on Agile practice usage and project data~\cite{Matthies2016}, this activity can be modified to present a more objective, data-based process view.

\paragraph{Quartering - Identify boring stories} assumes a collection of ``everything a team did'' in the last iteration.
The activity's description does not mention how this overview is achieved or how the data points are collected.
By brainstorming all their activities, this overview can be collaboratively reconstructed from the memories of participants.
Relying on project data could significantly speed up this (error-prone) method of data collection.
Dashboards featuring all interactions with the version control system by team members, e.g. using GitHub\footnote{\url{https://github.blog/changelog/2018-08-24-profile-activity-overview/}}, can present activity audits with minimal overhead, leaving more time in Retrospectives for discussion.
Furthermore, the goal of quartering is to identify boring stories.
While the ``boringness'' of a story/work item is, by definition, in the eye of the beholder, data from project issue trackers could provide an additional level of analysis:
Stories with no discussion that were closed rapidly, needing only a few commits by a single author, might be ideal candidates to be discussed for this Retrospective activity.

\paragraph{3 for 1} combines, as the name suggests, the assessments of meeting participants regarding three categories: iteration results, team communication, and mood.
Team members are asked to mark their spot in a coordinate plane using the axes ``satisfaction with iteration result'' and ``number of times we coordinated'' with an emoticon representing their mood.
While satisfaction with iteration results and mood are hard to gauge using project data, the frequency of communication within a team can be extracted from the team's employed communication tools.
As more communication moves to digital tools, such as chat or ticket systems, the wealth of information in this domain is steadily increasing~\cite{Stray2020}.
If a digital tool is used, the number of contacts and touch points between team members can be counted and quantified.
The input for one axis of the \emph{3 for 1} activity can therefore be automated or augmented with project data analyses.
Furthermore, variations of this exercise include varying the employed categories, such as replacing communication frequency with the frequency of pair programming~\cite{Kniberg2007} in the team.
Relying more heavily on project data analyses for this activity can simplify both data collection and substitution of employed categories.

\paragraph{Value Stream Mapping} attempts to create a \emph{value stream map} (VSM)~\cite{Fitzgerald2014,Kupiainen2015} of a team's process based on the perspective of a single user story.
While the details of the story might still be in participants' memories, gathering additional data, based on project artifacts, can provide additional context to improve the map's accuracy.
One of the main goals of a VSM is to identify delays, choke points, and bottlenecks in the process.
In a software development process, these are measurable using project data, e.g. by calculating the time it took from pushing code for a story until the code was reviewed or by assessing its \emph{lead time}~\cite{Ahmad2013}.
A more complete VSM can be generated by relying on these metrics, leading to improved subsequent analysis and improvement steps in a team.

\paragraph{Snow Mountain} uses the shape of the Scrum Burndown chart regarding a problematic iteration to draw an image that is used as a reflection prompt.
Using the metaphor of a snowy mountain ridge, meeting participants describe their perceptions of the iteration with kids sledging down the slopes.
The Burndown chart is a measurement tool for planning and monitoring of progress in Scrum teams~\cite{Scott2017}
They are based on the amount of work left to do versus remaining time during an iteration.
Depending on the team, the amount of outstanding work can be represented by time units,  story points or other effort measures (e.g. ``gummy bear''~\cite{Meyer14}).
If sophisticated project management software is used by the team and work items are entered into it with the required level of detail, burndown charts can be created and extracted from the project data\footnote{\url{https://support.atlassian.com/jira-software-cloud/docs/view-and-understand-the-burndown-chart/}}.
These digital images can then be printed or otherwise transformed into the snowy mountains required for the activity, without expending team members' time in creating them.

\section{Conclusion}
We present a review and analysis of current Retrospective activities, with a focus on the \emph{gather data} meeting phase.
We discuss the role of software project data, i.e. development artifacts produced by developers in their day-to-day work, within existing Retrospective meeting structures.
This type of data has previously been identified in the literature as an extremely valuable source of insight and actionable information \cite{Ortu2015,Guo2016}.
However, we show that the vast majority, i.e. 86\%, of activities explicitly proposed for the \emph{gather data} phase in a popular Retrospective agenda collection~\cite{Baldauf2018}, lack explicit connections to this software project data.
Of these data-gathering activities, many share a similar process of collecting participant perceptions and improvement ideas through structured prompts in the general form of \emph{start, stop, continue}.
Most current Retrospective activities rely on the perceptions of meeting participants as their sole inputs .
However, software project data, in particular requirements information or insights from version control systems, show promise as additional data sources for Retrospective techniques.
Integrating the principles of data-driven decision-making, based on project data, into Agile processes enables ``evidence-based decision making''~\cite{Fitzgerald2014} in Retrospective meetings.

These concepts are not foreign (or new) to Agile methods but seem to have fallen by the wayside recently.
The Scrum Guide states, ``Scrum is founded on empirical process control theory [...] knowledge comes from experience and making decisions based on what is known. [...]''~\cite{Schwaber2017}.
We argue that these concepts are still important, yet underused, in current implementations of Agile methods in general and Retrospectives in particular.

We identify four meeting activities in the Retromat collection that already explicitly take project data into consideration.
Of these, only two are listed for the \emph{gather data} phase of Retrospectives.
We then focus on employing software project data in additional activities to augment Retrospective meetings, decreasing manual efforts by Agile development teams, and process facilitators.

We propose modifications to five other activities, which are suited to take advantage of the process knowledge contained within project data.
These proposals present initial steps towards more evidence-based, data-informed decision making by participants of Retrospectives.


% \bibliographystyle{spmpsci}
% \bibliography{library}
\printbibliography

\clearpage
\section*{Appendix}
\begin{table}[!htbp]
    % \ra{1.2}
    \caption{List of activities extracted from the Retromat repository~\cite{Baldauf2018} for the \emph{gather data} phase of Retrospectives, as of Oct. 2020.}
    \label{table:activities}
    \begin{tabularx}{\textwidth}{lX}
        \toprule
        \textbf{\#} & \textbf{Name \& Activity Tagline} \\
        \midrule
        4 & \textit{Timeline}:  Write down significant events and order them chronologically \\
        5 & \textit{Analyze Stories}:  Walk through a team's stories and look for possible improvements \\
        6 & \textit{Like to like}:  Match quality cards to their own Start-Stop-Continue-proposals \\
        7 & \textit{Mad Sad Glad}:  Collect events of feeling mad, sad, or glad and find the sources \\
        19 & \textit{Speedboat/Sailboat}:  Analyze what forces push you forward and pull you back \\
        33 & \textit{Proud \& Sorry}:  What are team members proud or sorry about? \\
        35 & \textit{Agile Self-Assessment}:  Assess where you are standing with a checklist \\
        47 & \textit{Empty the Mailbox}:  Look at notes collected during the iteration \\
        51 & \textit{Lean Coffee}:  Use the Lean Coffee format for a focused discussion of the top topics \\
        54 & \textit{Story Oscars}:  The team nominates stories for awards and reflects on the winners \\
        62 & \textit{Expectations}:  What can others expect of you? What can you expect of them? \\
        64 & \textit{Quartering}:  Categorize stories in 2 dimensions to identify boring ones \\
        65 & \textit{Appreciative Inquiry}:  Lift everyone's spirit with positive questions \\
        75 & \textit{Writing the Unspeakable}:  Write down what you can never ever say out loud \\
        78 & \textit{4 Ls}:  Explore what people loved, learned, lacked and longed for individually \\
        79 & \textit{Value Stream Mapping}:  Draw a value stream map of your iteration process \\
        80 & \textit{Repeat \& Avoid}:  Brainstorm what to repeat and what behaviours to avoid \\
        86 & \textit{Lines of Communication}:  Visualize information flows in, out and around the team \\
        87 & \textit{Meeting Satisfaction Histogram}:  Create a histogram on how well ritual meetings went during the iteration \\
        89 & \textit{Retro Wedding}:  Collect examples for something old, new, borrowed and blue \\
        93 & \textit{Tell a Story with Shaping Words}:  Each participant tells a story about the last iteration that contains certain words \\
        97 & \textit{\#tweetmysprint}:  Produce the team's twitter timeline for the iteration \\
        98 & \textit{Laundry Day}:  Which things are clear \& feel good and which feel vague \& implicit? \\
        110 & \textit{Movie Critic}:  Imagine your last iteration was a movie and write a review about it \\
        116 & \textit{Genie in a Bottle}:  Playfully explore unmet needs \\
        119 & \textit{Hit the Headlines}:  Which sprint events were newsworthy? \\
        121 & \textit{The Good, the Bad, and the Ugly}:  Collect what team members perceived as good, bad and non-optimal \\
        123 & \textit{Find your Focus Principle}:  Discuss the 12 agile principles \& pick one to work on \\
        126 & \textit{I like, I wish}:  Give positive, as well as non-threatening, constructive feedback \\
        127 & \textit{Delay Display}:  What's the current delay? And where are we going again? \\
        128 & \textit{Learning Wish List}:  Create a list of learning objectives for the team \\
        133 & \textit{Tell me something I don't know}:  Reveal hidden knowledge with a game show \\
        135 & \textit{Avoid Waste}:  Tackle the 7 Wastes of Software Development \\
        137 & \textit{Dare, Care, Share}:  Collect topics in three categories: 'Dare', 'Care' and 'Share' \\
        139 & \textit{Room Service}:  Take a look at the team room: Does it help or hinder? \\
        \bottomrule
    \end{tabularx}
\end{table}

\end{document}